\documentclass[aps,prl,twocolumn,amsmath,amssymb,floatfix,superscriptaddress]{revtex4}
\usepackage{graphicx,subfigure,psfrag}
\newcommand{\id}{{\mathbf 1}}

\allowdisplaybreaks[4]
\begin{document}
\title{Non-Abelian statistics in the interference noise of the Moore-Read quantum Hall state}

\author{Eddy Ardonne}
\affiliation{Center for the Physics of Information, California Institute of Technology, Pasadena, CA 91125, USA}
\affiliation{Microsoft Station Q, University of California, Santa Barbara, CA 93106, USA}
\author{Eun-Ah Kim}
\affiliation{Stanford Institute for Theoretical Physics and Department of Physics, Stanford University, Stanford, CA 94305, USA}
\begin{abstract}
We propose noise oscillation measurements in a double point contact, accessible with current technology, to seek for 
a signature of the non-abelian nature of the $\nu\!=\!5/2$ quantum Hall state. 
Calculating the voltage and temperature dependence of the current and noise oscillations, we predict the non-abelian nature to materialize through a multiplicity of the possible outcomes: two qualitatively different frequency dependences of the nonzero interference noise.
Comparison between our predictions for the Moore-Read state with experiments on $\nu\!=\!5/2$ will serve as a much needed test for the nature of the $\nu\!=\!5/2$ quantum Hall state.
\end{abstract}
\maketitle
Non-abelian quantum Hall (QH) states, such as the Moore-Read (MR) QH state\cite{mr} 
are considered to be the most promising route\cite{pt} to fault tolerant topological quantum computation\cite{kitaev}.
The possibility of the $\nu\!=\!5/2$ QH plateau\cite{willet} being the MR state\cite{gww,num52} attracted interests from a wide range of fields:  from string theory to solid state physics.
A configuration of many non-abelian excitations, such as MR quasiholes/particles (qh's/ qp's),  is associated with a set of {\it degenerate} states. An exchange of two such excitations amounts to a rotation in the degenerate state space:
the most exotic form of statistics allowed in two space dimensions. For the MR state, the $2n$ qh state is $2^{n-1}$ fold degenerate\cite{nw} and 4-qh's can form a single quantum bit (qubit).  This notion is at the heart of current enthusiasm for MR state, from both fundamental science and application oriented view. However, non-abelian statistics has not been observed to date.

There are proposals for detecting non-abelian statistics of MR state by exploiting the braiding properties of underlying Chern-Simons theories\cite{dfn,sh1,bks, fk,hc}. Effects of non-abelian statistics on the non-linear transport of a single point contact has also been predicted\cite{bn,ffn}.
While a signature of non-abelian statistics is yet to be observed, 
a recent experiment \cite{cmarcus} demonstrated the feasibility of 
a $\nu\!=\!5/2$ single point contact (PC), whose qualitative tunneling characteristics are those of the MR edge state. Thus, the edge states can be used as probes~\cite{wen95}  of the exotic topological order 
associated with the $\nu\!=\!5/2$ state.

In this letter, we propose feasible noise measurements in a
double PC interferometer and give a detailed prediction on clear, qualitative signatures of the non-abelian
statistics at finite temperature and voltage. 
A noise spectrum  is a powerful probe for the nature of excitations since it is 
determined by the dynamical properties  containing information about the excited states.
It is well known (see Ref\cite{lesovik94}) that the noise spectrum of an electronic system in an appropriate geometry can contain statistics-dependent features that are not contained in the dc conductance. 
It is natural to expect that the noise spectrum of strongly interacting systems, such as quantum Hall liquids, should exhibit an even richer behavior such as the one found in the case of the Abelian FQH states (Ref.\cite{kim:176402}), and for the non-Abelian case in this paper.
 Here we focus on the double PC setup for two reasons: 
i) It is the only interferometer that has been experimentally realized in (abelian) fractional QH states\cite{goldman} which was subsequently analyzed theoretically\cite{kim-goldman}.
 ii) By attaching leads to the edge states of this setup it is possible to realize the physical situation of 
 the four qh state which is the simplest state in which the consequences of their non-Abelian statistics become directly observable.
 We examine the oscillatory part of the noise as a cross current fluctuation and present the leading order perturbation theory result. Our results apply to both Abelian and non-Abelian cases.
In addition, we provide an interpretation of the `even-odd effect' \cite{sh1,bks} in the context of the edge state theory.

\begin{figure}[t] 
   \includegraphics[width=.35\textwidth]{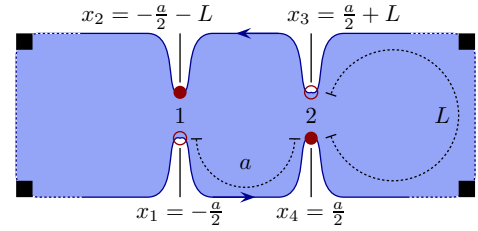} 
   \caption{The double point contact setup, indicating the four positions $x_i$, $i=1,\ldots,4$
   associated with the point contacts $1$ and $2$, in terms of the chiral abscissa coordinate $x$
   (defined modulo $2a+2L$) for the chiral edge.}
    \label{fig:setup}
\end{figure}
{\it Double PC interferometer} --
The double PC setup was first proposed as a testbed for abelian fractional statistics\cite{cfksw} 
and there have been discussions on using the setup to detect non-abelian
statistics~\cite{fntw,dfn,sh1,bks}.
It was first pointed out in Ref.\cite{fntw} that the interference  between two different paths of {\it adiabatic transport}  surrounding a region with localized qh's can be used to measure the associated non-Abelian braiding. While this picture provides conceptual intuition,  
an explicit calculation in terms of the edge theory is still needed.

The edge state theory relevant for the low energy dynamics of  {\it gapless} edge excitations 
of the MR state consists of two parts:
the standard free chiral boson $\varphi_c$ describing the charge modes \cite{wen} with
$\mathcal{L}_c\!=\!1/(2\pi)\partial_x\varphi_c(\partial_t \!+\!v\partial_x)\varphi_c$, where $v$ is the edge mode velocity, and an
additional charge neutral part: the chiral Ising conformal field theory (CFT), with a free Majorana field $\psi$ and the spin-field $\sigma$~\cite{wen95}. The non-abelian nature of the Ising CFT is encoded in the fusion rule $\sigma\! \times\! \sigma \!=\! \id \!+\! \psi$
, which makes the correlator of multiple $\sigma$'s to form multi-dimensional conformal blocks\cite{bpz}. 
The qh creation operator
$\sigma e^{i/\sqrt{8} \varphi}(z)$ ($z\!\equiv\! i(vt\!-\!x)$) 
is the most relevant operator in the renormalization group sense. 

\begin{figure}[ht] 
   \includegraphics[width=.3\textwidth]{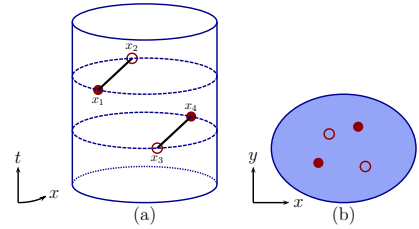}
   \caption{Four points marked in (a) $1\!+\!1$ D `event' space and in (b) $2\!+\!0$ D space. CFT correlators associated with marked points are interpreted as 
   (a) `vacuum expectation values' associated with current carrying gapless edge modes (the black lines represent the Wilson lines, see text), (b) `wave functions' associated with excitation configurations with finite energy.} 
   \label{fig:example} 
\end{figure}

The manifestly relativistic nature of the edge CFT in $1\!+\!1$ D  reflects the general covariance (or topological invariance) of the underlying $2\!+\!1$ D Chern-Simons theory~\cite{witten}, as a low energy effective field theory of the quantum Hall liquid~\cite{fntw,fns}. An edge qh operator $\sigma e^{i\sqrt{8}\varphi}(z)$ `marks' a point on the $1\!+\!1$ D surface, which corresponds to an end point of a Wilson line in the $2\!+\!1$ D Chern-Simons theory bounded by the surface (see Fig.~\ref{fig:example}(a)). 
Witten first showed that the multi-dimensionality of CFT correlators represented by the fusion rules reflects the non-abelian statistics of the corresponding qh/qp's represented by the associated Wilson lines. 
Moore and Read~\cite{mr} proposed to interpret CFT correlators to represent many body wave functions for  quantum Hall states, now with the complex coordinate $z\!\equiv\!x\!+\!iy$ defined in $2\!+\!0$ D (see Fig.~\ref{fig:example}(b)). The MR state wave function so constructed  from the Ising CFT, became a candidate description of the $\nu\!=\!5/2$ state \cite{gww}.   
Nayak and Wilczek~\cite{nw} further demonstrated the non-abelian nature of the four-qh wave function through explicit  exchange operations.
While the wave function gives a clear physical picture of the nature of the state, it in itself is not a measurable quantity. On the other hand, the edge CFT can bridge between the theoretical structure and measurements.

We start by observing that a  double PC 
allows one to access the four point $\sigma$ correlator in the $1\!+\!1$ D edge CFT
(Fig.~\ref{fig:example}(a)).
A quantum mechanical tunneling event annihilates a particle at one side of a point contact while creating one on the other side.
Tunneling response  
naturally calls for contributions from 
four point functions with different ordering of $\sigma$ operators in this event space, at leading order.  Hence, the tunneling response incorporates effects of exchange in the $1\!+\!1$ D event space.

{\it Perturbative calculation} --
We model the double PC setup with separation $a$ between two PC's using a single abscissa coordinate $x$, defined modulo $2a+2L$ (see figure \ref{fig:setup}) and time $t$, which parametrize a cylinder.
By taking the limit $L\rightarrow\infty$ at the end of the calculation, we take the effect of the leads into account properly without allowing the edge current to go around the whole sample\cite{fn}. 
While it is typical to combine two chiral modes to form a single nonchiral mode \cite{cfksw,ffn}, it is not possible to take such approach for a MR multi PC setup without losing the information about its intricate topological structure.
Our procedure allows us to  describe the system using a single chiral edge mode; we checked it against the non-chiral mode approach
in the abelian case \cite{ardkim}.

The operators which tunnel a qp 
at PC's $1$ and $2$ are \vspace{-2mm}
\begin{multline}
\hat{V}_1 (t) = \sigma (x_1,t) \sigma (x_2,t) e^{i/\sqrt{8}\varphi_c(x_1,t)}e^{-i/\sqrt{8}\varphi_c(x_2,t)} \\
\hat{V}_2 (t) = \sigma (x_3,t) \sigma (x_4,t) e^{i/\sqrt{8}\varphi_c(x_3,t)}e^{-i/\sqrt{8}\varphi_c(x_4,t)} \ ,
\end{multline}
which accounts for creation and annihilation of qp's  on opposite edges at equal time
(we note that $\sigma$ is self dual).
The appropriate tunneling hamiltonian and the current operator are then
$\hat{H}_{\rm tun} (t) = \sum_j \Gamma_j (t) \hat{V}_j (t) + {\rm h.c.}$ and 
$\hat{I} (t) = i e^* \sum_j \Gamma_j (t) \hat{V}_j (t) + {\rm h.c.}$\cite{PhysRevB.51.2363}. 
Here the time dependent tunneling strength is given by
$\Gamma_j (t) = \Gamma_j e^{i\omega_0 t}$, with $\omega_0= \tfrac{e^*V}{\hbar}$ the
Josephson frequency and $e^*=e/4$ the charge of the tunneling quasi-particle.
In a magnetic field, the Aharonov-Bohm phase acquired by tunneling qp's can be effectively
incorporated through a 
flux dependent relative phase between two tunneling amplitudes as
$\Gamma_1\Gamma_2^*\!=\! |\Gamma_1\Gamma_2|e^{i\phi/\Phi_0}$\cite{cfksw}.
We will assume that the tunneling is sufficiently weak at finite temperature and voltage and 
that the  lowest Landau level is inert. We only consider tunneling of the most relevant quasi-holes.

We now calculate the average steady state current
\begin{equation}
\langle \hat{I} \rangle = -i \int_{-\infty}^t dt' \langle[\hat{I} (t), \hat{H}_{\rm tun} (t')]\rangle, \label{eq:I}
\end{equation}
which in general is a highly non-linear function of $V$ and $T$ for finite separation $a$, and the non-equilibrium noise
\begin{equation}
S(\omega) = \frac{1}{2} \int_{-\infty}^{\infty} dt' e^{i \omega t'}\langle \{ \hat{I} (t),\hat{I}(t') \} \rangle, 
\label{eq:S}
\end{equation}
which we define as the usual two-point correlations involving the operator $\hat{I}$\cite{PhysRevB.51.2363}.
Notice that current, which is a {\it causal} response, invloves a commutator; while the noise, which is a {\it fluctuation}, involves an anti-commutator. This basic fact, when applied to $1+1$D edge tunneling transport, has non-trivial consequences, both in that Eqs.~(\ref{eq:I}-\ref{eq:S}) require exchange in the event space and that one is restricted by causality while the other is not.  It is amusing that 
both Eq.~\eqref{eq:I} and \eqref{eq:S} explicitly depend on four-$\sigma$ correlator at lowest order in $\Gamma$, which can take two possible values due to the non-Abelian nature of the $\sigma$ operators. We label these two possibilities by $p=0,1$.  

\begin{figure*}[bht]
   	\includegraphics[width=.70\textwidth]{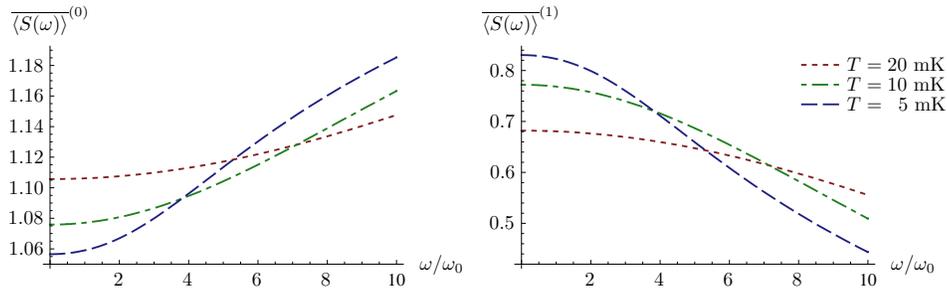}
   \caption{The (dimensionless) amplitude of the noise oscillations
   $\overline{\langle S(\omega)\rangle}^{(p)}$ in two states (0) (left) and (1) (right) for
   $a\!=\!5\mu$m, $V\!=\!.38 \mu V$ ($\omega_0 \!= \!100$ MHz),   $v \!=\! 1.10^7 m/s$, for various temperatures.}\vspace{-3mm}
   \label{fig:noise-amp}
\end{figure*}
To leading order, the current and noise are ($e^*\!=\!1$)
\begin{align}
{\langle\hat{I}\rangle^{(p)}} \!&\equiv \langle\hat{I}\rangle^{(p)}_d + \cos (\tfrac{\phi}{\Phi_0}) \langle\hat{I}\rangle^{(p)}_{\rm osc} =\label{eq:curcor}\\
 \Re\!\int_{0}^{\infty} \!\!\!\!dt\!\!  \sum_{j,k=1}^2& \Gamma_j\Gamma_k^*
\Bigl[ e^{-i\omega_0 t}
\bigl( \langle \hat{V}_j\hat{V}_k^\dagger\rangle^{(p)} (t) 
\!-\! \langle \hat{V}_k^\dagger \hat{V}_j \rangle^{(p)} (t)\bigr) \Bigr]\nonumber\\
{\langle S(\omega) \rangle^{(p)}}\! &
\equiv\left( \langle S(\omega) \rangle^{(p)}_d + \cos(\tfrac{\phi}{\Phi_0}) \langle S(\omega) \rangle^{(p)}_{\rm osc}\right)=\label{eq:noisecor}\\
 \Re\!\int_{-\infty}^{\infty}\!\!\!\!dt\!  \sum_{j,k=1}^2&
\Gamma_j\Gamma_k^*
\Bigl[ e^{i(\omega-\omega_0) t}
\bigl( \langle \hat{V}_j\hat{V}_k^\dagger\rangle^{(p)} (t) 
\!+\! \langle \hat{V}_k^\dagger \hat{V}_j \rangle^{(p)} (t)\bigr) \Bigr]\nonumber
\end{align}
with the Aharonov-Bohm oscillatory parts which require coherence between two PC's, and
the direct parts which only involve a single PC.
The correlators $\langle \hat{V}_j\hat{V}_k^\dagger\rangle^{(p)} (t)$'s can be calculated 
 in terms of $s(x,t)\!\equiv\! i \sinh (\pi T(x+t))$ using the standard conformal mapping: \vspace{-2mm}
\begin{widetext}
\begin{equation}
\langle \hat{V}_1\hat{V}_2^\dagger\rangle^{(p)} (t) =
\sqrt{\pi T/2} \frac{(-1)^p s(a+L,t)^{1/4}s(-a-L,t)^{1/4}}{s(-L,t)^{1/4}s(-a-2L,t)^{1/4}s(a,t)^{1/4}s(-a,t)^{1/4}}\
\sqrt{1+(-1)^{p}\sqrt{\frac{s(a,t)s(-a,t)}{s(a+L,t)s(-a-L,t)}}} \ . 
\label{eq:V}
\end{equation} \vspace{-2mm}
 \end{widetext}
Here, the phase is determined by requiring the current to flow in the right direction. We used the chiral boson correlator combined with the four $\sigma$ correlator 
\begin{multline}
\langle \sigma(z_1)\sigma(z_2)\sigma(z_3)\sigma(z_4)\rangle^{(p)} =
\frac{1}{\sqrt{2}} (z_1-z_2)^{-\frac{1}{8}}(z_3-z_4)^{-\frac{1}{8}}\\ \times
(1-\xi)^{-1/8}\sqrt{1+(-1)^p\sqrt{1-\xi}} \ ,
\label{eq:4sigma}
\end{multline}
with the cross-ratio  $\xi=\frac{(z_1-z_2)(z_3-z_4)}{(z_1-z_4)(z_3-z_2)}$\cite{bpz}.

{\it Hearing the non-abelian statistics} --
Rather unexpectedly, we find that the channel dependence,  a hallmark of non-abelian statistics,  shows up in the noise but not in the current. This is due to an intriguing interplay between the inherently relativistic nature of the edge state theory and the causal nature of current as a response. 
The evaluation of the tunneling current and the noise of Eqs.~(\ref{eq:curcor}-\ref{eq:noisecor}) requires combining four different terms of the type
$\langle \hat{V}_j\hat{V}_k^\dagger\rangle^{(p)} (t)$.
Explicitly exchanging these operators, with much attention to branch cuts and taking the limit
$L\rightarrow\infty$ afterwards, we find for the Aharonov-Bohm oscillation amplitude of the current and the noise \cite{ardkim} 
\begin{align}
\label{eq:currentosc}
& \langle\hat{I}\rangle^{(p)}_{\rm osc} = 4e^* \sqrt{\pi T}|\Gamma_1\Gamma_2|
\times \\ \nonumber & 
\int_{a}^{\infty} dt \frac{\sin(\omega_0 t)}{\sinh(\pi T(t-a))^{1/4}\sinh(\pi T(t+a))^{1/4}} \\
\label{eq:noiseosc}
& \langle S(\omega) \rangle^{(p)}_{\rm osc} = 4 (e^*)^2 \sqrt{\pi T} |\Gamma_1\Gamma_2|
\times \nonumber \\ &
\Bigl(
\int_{a}^{\infty} dt \frac{\cos((\omega+\omega_0)t)+\cos((\omega-\omega_0)t)}
{\sinh(\pi T(t-a))^{1/4}\sinh(\pi T(t+a))^{1/4}} + \\ & (-1)^p 
\int_{0}^{a} dt \frac{\sqrt{2}(\cos((\omega+\omega_0)t)+\cos((\omega-\omega_0)t))}
{\sinh(\pi T(t-a))^{1/4}\sinh(\pi T(t+a))^{1/4}} \Bigr) \ . \nonumber
\end{align}
The direct, single PC contributions which yields the shot noise results ($S(0)=e^*\langle I\rangle$) can be obtained from Eqs.~(\ref{eq:currentosc}-\ref{eq:noiseosc}) by taking the limit $a\rightarrow 0$.  The above results can be generalized to other FQH states in a straight forward manner. Note that for (Abelian) Laughlin states at filling $\nu=1/m$,  the exponents change from $1/4$ to $1/m$ and only the $p\!=\!0$ state is possible.

The state $(p)$ dependence only appears in the second term of the noise oscillation Eq.~\eqref{eq:noisecor}, which is absent in the single PC limit. 
In order to understand this we note that the `light cone' $|t|\!=\!a$  (here the speed of light is the edge mode speed $v\!\equiv\!1$) divides causally connected (time-like separated) region $|t|\!>\!a$ from the space-like separated region $|t|\!<\!a$, for the correlator Eq.~\eqref{eq:V}.
Due to the branch cut structure, the correlators behave differently under exchange of operators in these two event space regions, and we find the channel dependence to vanish in any causally connected regions ($|t|\!>\!a$)~\cite{ardkim}. 
Hence the current, which is a causal response (see Eq.~\eqref{eq:I}), is state independent. In contrast, the noise, which is a fluctuation unrestricted by causality (see ~Eq.~\eqref{eq:S}), will display state dependence when the space-like separated contribution is significant.

Fig.~\ref{fig:noise-amp} shows a clear, qualitative difference in the frequency
dependence of the noise oscillations in the two states: $p\!=\!0$ and $p\!=\!1$. Here we plotted the dimensionless noise oscillation amplitude  $ \overline{\langle{S}(\omega)\rangle}^{(p)}\!\equiv\! 
\langle S(\omega) \rangle^{(p)}_{\rm osc}/\langle S(\omega)\rangle_d$,
 (assuming $|\Gamma_1|\!=\!|\Gamma_2|$), for different temperatures and parameters within reach of current technology.  We find that there is optimal range for the distance $a$. If $a$ is too small, the system reaches single PC limit without state dependence. On the other hand, if $a$ becomes comparable to the thermal decoherence scale set by $T$, interference features get washed out as $\exp(-a T)$~\cite{ardkim}. 
The qualitative difference in the two states traces back to the fact that the contributions from two event-space regions are added for state $(0)$ while they are to be subtracted for state $(1)$
(see Eq.~\eqref{eq:noiseosc}).
This relative negative sign for the state $(1)$ reflects the hidden majorana fermionic character of this state which is symbolically represented in the second term of the fusion rule, $\sigma\times\sigma={\mathbf 1}+\psi$. 
Only in state $(1)$ the rearrangement of $\sigma$'s needed in Eqs.~(\ref{eq:curcor}-\ref{eq:noisecor})
effectively exchanges two majorana fermions. 
This fermionic nature results in the decreasing concave frequency dependence. 

\begin{figure}[t]
\psfrag{a}{(0)}
\psfrag{b}{(1)}
\psfrag{fsqrt2}{\small$=\!-\sqrt{2}$}
\includegraphics[width=.40\textwidth]{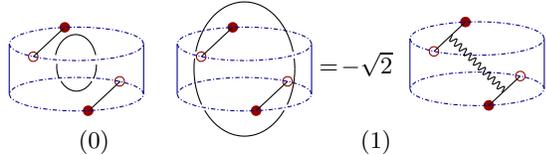}
\caption{ Two distinct states of double PC. (left) A Wilson loop which can be shrunk to a point. (right) A Wilson loop looping around both Wilson lines which is equivalent to two Wilson lines exchanging a $\psi$ up to a factor of $-\sqrt{2}$.}\vspace{-5mm}
\label{channels}
\end{figure}
{\it Effect of localized qh's.} --
The distinction between these two equally allowed states is the configuration of bulk quasi holes. 
We depicted  two topologically distinct Wilson line configurations corresponding to states $(0)$ and $(1)$ in Fig.\ref{channels}.
The underlying braid properties of the Wilson lines \cite{mooreseiberg,ardkim} for the two configurations of Fig.~\ref{channels} allows the interference noise to access the two state nature (and hence the essence of the non-abelian statistics) of the MR state.
The observed result depends on the state of the system. For instance, if the bulk quasiparticles are in a mixed state, the measured noise will be a linear combination of the results for state (0) and (1). 
This is in contrast to the abelian case which can only be in the pure state (0) since in this case the state is unique\cite{ardkim}. Hence, even a maximally mixed state will have a signature that distinguishes the non-Abelian state with two dimensional Hilbert space from an Abelian state.

The Aharonov-Bohm oscillations vanish for both the noise and the current when the Wilson line of a qp-qh pair in the bulk loops around only {\it one} of the Wilson lines associated with the tunneling.
In this case, one pair of $\sigma$'s fuses to $\psi$ while the other fuses to {\bf 1}, 
and thus the correlator Eq.~\eqref{eq:4sigma} vanishes \cite{ardkim}.
This is the edge-theory interpretation of the even-odd effect \cite{sh1,bks}.
We have shown that there are two distinct possibilities, what we called $(0)$ and $(1)$, within the non-vanishing `even' case which is evident in the interference noise. This provides an alternative way of looking for the signature of non-Abelian statistics, which can easily be generalized for other non-Abelian states. 

{\it Conclusion}\;\;
We perturbatively calculated the tunneling current and noise of a double PC interferometer in the MR quantum Hall state using the associated edge state theory. This setup provides direct experimental access to the four-$\sigma$ correlator which describes two topologically distinct states. Exploiting the fact that the measurable quantities naturally involve exchange  in the {\it event space}, we find that the Aharonov-Bohm oscillatory noise can be used to ``hear'' a clear signature of non-abelian statistics.
We predict a qualitative difference in the low frequency behavior of the oscillatory noise between the two states. Our detailed predictions for the voltage and temperature dependence can be compared with future measurements. 
Due to the non-local entanglement between bulk and edge qh's, which is tied to the non-Abelian nature, the preparation of a system in a pure state of any of the topologically distinct possibilities considered here or in Refs.\cite{sh1, bks} requires the control of pinned bulk qh's. 
The problem of how to effectively control the state is an important and open question of direct relevance to experiments on non-Abelian interferometers.

\noindent
{\bf Acknowledgments}:
We thank P.\@Bonderson, C.\@Chamon, S.B.\@Chung, L.\@Fidkowski, E.\@Fradkin, M.\@Freedman, C.\@Nayak, S.\@Shenkar, J.\@Slingerland, K.\@Shtengel, Z.\@Wang for illuminating discussions.  
EAK was supported by the Stanford Institute for Theoretical Physics and in part by the Microsoft Station Q.

\vspace{-4mm }

\end{document}